\documentclass[5p,twocolumn]{elsarticle}
\usepackage{lineno,hyperref}

\usepackage{graphics,graphicx}
\usepackage{amsmath,amssymb,amsfonts,amsthm}

\allowdisplaybreaks[4]

\begin{document}

\begin{frontmatter}
\title{Nuclear probes of an out-of-equilibrium plasma at the highest compression}
\author[SARI,SINAP]{G. Zhang}
\ead{zhangguoqiang@sinap.ac.cn}
\author[IMU]{M. Huang}
\author[TMAU,INFN]{A. Bonasera}
\ead{abonasera@comp.tamu.edu}
\author[FDU,SINAP,UCAS]{Y.G. Ma}
\ead{mayugang@cashq.ac.cn}
\author[SIOM,SNU]{B.F. Shen}
\ead{bfshen@mail.shcnc.ac.cn}
\author[SARI,SINAP]{H.W. Wang}
\author[SIOM]{W.P. Wang}
\author[SIOM]{J.C. Xu}
\author[SARI,SINAP]{G.T. Fan}
\author[SINAP]{H.J. Fu}
\author[SINAP]{H. Xue}
\author[SNU1]{H. Zheng}
\author[SARI,SINAP]{L.X. Liu}
\author[IMU]{S. Zhang}
\author[SINAP]{W.J. Li}
\author[SARI,SINAP]{X.G. Cao}
\author[SINAP]{X.G. Deng}
\author[SINAP]{X.Y. Li}
\author[SINAP]{Y.C. Liu}
\author[SIOM]{Y. Yu}
\author[SINAP]{Y. Zhang}
\author[FDU,SJTU]{C.B. Fu}
\author[SJTU]{X.P. Zhang}

\address[SARI] {Shanghai Advanced Research Institute, Chinese Academy of Sciences, Shanghai 201210, China.}
\address[SINAP] { Shanghai Institute Applied Physics, Chinese Academy of Sciences, Shanghai 201800, China.}
\address[IMU]{College of Physics and Electronics information, Inner Mongolia University for Nationalities, Tongliao, 028000, China.}
\address[TMAU]{Cyclotron Institute, Texas A\&M University, College Station, Texas 77843, USA.}
\address[INFN]{Laboratori Nazionali del Sud, INFN, via Santa Sofia, 62, 95123 Catania, Italy.}
\address[FDU] {Key Laboratory of Nuclear Physics and Ion-beam Application (MOE), Institute of Modern Physics, Fudan University, Shanghai 200433, China}
\address[SIOM]{ State Key Laboratory of High Field Laser Physics, Shanghai Institute of Optics and Fine Mechanics,Chinese Academy of Sciences, Shanghai 201800, China.}
\address[SNU]{Department of Physics, Shanghai Normal University, Shanghai 200234, China}
\address[UCAS]{ University of the Chinese Academy of Sciences, Beijing 100080, China.}
\address[SNU1]{ School of Physics and Information Technology, Shaanxi Normal University Xi'an 710119, China.}
 \address[SJTU]{School of Physics and Astronomy, Shanghai Jiao Tong University, Shanghai 200240, China.}

\begin{abstract}
We report the highest compression reached in laboratory plasmas using eight laser beams, E$_{laser}$$\approx$12 kJ, $\tau_{laser}$=2 ns in third harmonic on a CD$_2$ target at the ShenGuang-II Upgrade (SGII-Up) facility in Shanghai, China. We estimate the deuterium density $\rho_D$= 2.0 $\pm$ 0.9 kg/cm$^{3}$, and the average kinetic energy of the plasma ions less than 1 keV.  The highest reached areal density $\Lambda \rho_{D}$=4.8 $\pm$ 1.5 g/cm$^{2}$ was obtained from the measured ratio of the sequential ternary fusion reactions (dd$\rightarrow$t+p and t+d$\rightarrow$$\alpha$+n) and the two body reaction fusions (dd$\rightarrow$$^3$He+n). At such high densities, sequential ternary and also quaternary nuclear reactions become important as well (i.e. n(14.1 MeV) + $^{12}$C $\rightarrow$ n'+$^{12}$C* etc.) resulting in a shift of the neutron (and proton) kinetic energies from their birth values. The Down Scatter Ratio (DSR-quaternary nuclear reactions) method, i.e. the ratio of the 10-12MeV neutrons divided by the total number of 14.1MeV neutrons produced, confirms the high densities reported above. The estimated lifetime of the highly compressed plasma is 52 $\pm$ 9 ps, much smaller than the lasers pulse duration. 
\end{abstract}

\end{frontmatter}






The understanding of the fascinating supernovae explosions\cite{Burbidge1957,Wallerstein1997,Woosley2002,Remington1999} as well as the social requirement of clean, cheap and easily available energy\cite{Clery2014} obtainable from nuclear fusion power plants require the microscopic understanding of the nuclear reactions in plasmas. Especially for hot and very dense plasmas, such as in the interior of a star, or in a highly compressed nuclear fuel, it is crucial to know the probability of fusion and the range of the ions at the density and temperature of the system. Fusion reactions are usually measured in beam-target experiments and are reliable for relatively large beam energies. At low beam energies, the probabilities are extrapolated from higher energies through direct\cite{Brown1990,Krauss1987,Mukhamedzhanov2007,Leonard2006,Angulo1999,Xu2013} and indirect methods\cite{Angulo1999, Xu2013, Li2015, Tribble2014, Pizzone2013}, such as  the Trojan Horse method (THM), the asymptotic normalization coefficient (ANC) and  the Coulomb dissociation method (CD), but all these methods do not guarantee that in the hot and dense plasma the fusion reaction process is not influenced by the motion of the electrons and the other ions. Furthermore, the range of the ions in the plasma is quite different from the range in a cold target. We can express the probability of fusion as:
\begin{equation}
 \prod = 1- e^{-\Lambda/\lambda}=1-e^{-\Lambda\rho\sigma}\cong\Lambda\rho\sigma
\label{eq1:prod}
\end{equation}
which contains all the ingredients needed for understanding the dynamics of fusion in the plasma.  $\Lambda$ is the range of the ion in the plasma, i.e. the distance travelled by the ion before loosing its kinetic energy; it is relatively well known and understood in cold targets as the slowing down of the ions due to electromagnetic interactions (plus nuclear processes at high energy)\cite{Ziegler2010}.  The nuclear mean free path $\lambda$ is expressed as $\lambda=1/(\rho \sigma)$, where $\rho$ stands for the deuterium number density of the plasma. $\sigma$ is the nuclear fusion reaction cross-section depending on the center of mass (C.M.) energy of the colliding ions and it is usually measured in beam-(cold) targets experiments\cite{Angulo1999, Xu2013}. New methods to measure these quantities directly in the plasma have been recently investigated using a petawatt laser impinging on a cluster target. Fusion cross sections for d+$^3$He\cite{Barbui2013}, dd\cite{Lattuada2016} and the range\cite{Zhang2017} have been measured also in the cases where the system is prepared near the critical point for a liquid gas phase transition\cite{Quevedo2018}. While the fusion cross sections have been found in reasonable agreement with accelerator experiments, the range showed some dependence on the cluster distribution, which in turn depends on the equation of state of the system\cite{Zhang2017,Quevedo2018}. Consequently, according to eq.(\ref{eq1:prod}), the number of fusions is influenced as well. The typical ion density in these experiments is of the order of 10$^{18} $ cm$^{-3}$, the target size around 0.5 cm and the temperature of the plasma tens of keV. Strong non-equilibrium effects did not prevent a good understanding of the plasma dynamics and a precise measurement of the ingredients entering eq.(\ref{eq1:prod}). Careful methods might be devised to keep the system as close as possible to equilibrium for instance using several laser beams to compress and heat a target\cite{Hurricane2014}. Nevertheless, important out of equilibrium effects might still be present\cite{Zheng2013,LePape_prl2018,Casey_prl2012} and prevent the optimization of nuclear fusion reactions towards a prototype of a nuclear reactor. Stimulated by these considerations we decided not to fight non-equilibrium effects but rather enhance them, i.e. study plasmas highly compressed and completely out of equilibrium.  A scheme for a colliding beam fusion reactor has been proposed as well\cite{Rostoker1997}.  Recent experiments have also shown a substantial increase in the number of fusions using the Target Normal Sheath Acceleration (TNSA) mechanism\cite{Wilks2001}, which utilizes short pulse lasers to accelerate light ions to impinge on a cold target\cite{Krygier2015, Morrison2012} or on a plasma\cite{Labaune2013}. In previous experiments at the ABC laser facility\cite{Barbarino_phd2015, Perrotta2019} these features have been explored using two laser beams strongly focalized on a flat and thin target from opposite directions. Microscopic simulations for out of equilibrium systems\cite{Bonasera2004} suggest that opportunely choosing the target size and material, it is possible to reach high densities where catastrophic nuclear process might occur.  Recent models have been proposed, also with the geometry discussed below \cite{Csernai2019,Csernai2018,Csernai2015} , which  could serve as inspiration for future experiments. However, the targets, material composition and laser features require some effort and considerable fundings. Thus we hope that the results discussed here might stimulate some theoretical analysis to test the model assumption and give further impetus to this line of research. A problem might occur if the system is also not locally neutral (a large number of electrons are extracted from the target at the beginning of the laser-target interaction) which might influence the fusion probability and in particular decrease it as compared to beam-cold target experiments \cite{Barbarino_phd2015}. These non-equilibrium effects might not be properly included in hydrodynamic simulations thus reducing their predictive power. Strong (anti) screening effects might be at play and hinder the nuclear reaction rates as well\cite{Huang2018}. 

\begin{figure}[h]
\centering
\includegraphics[width=0.45\textwidth]{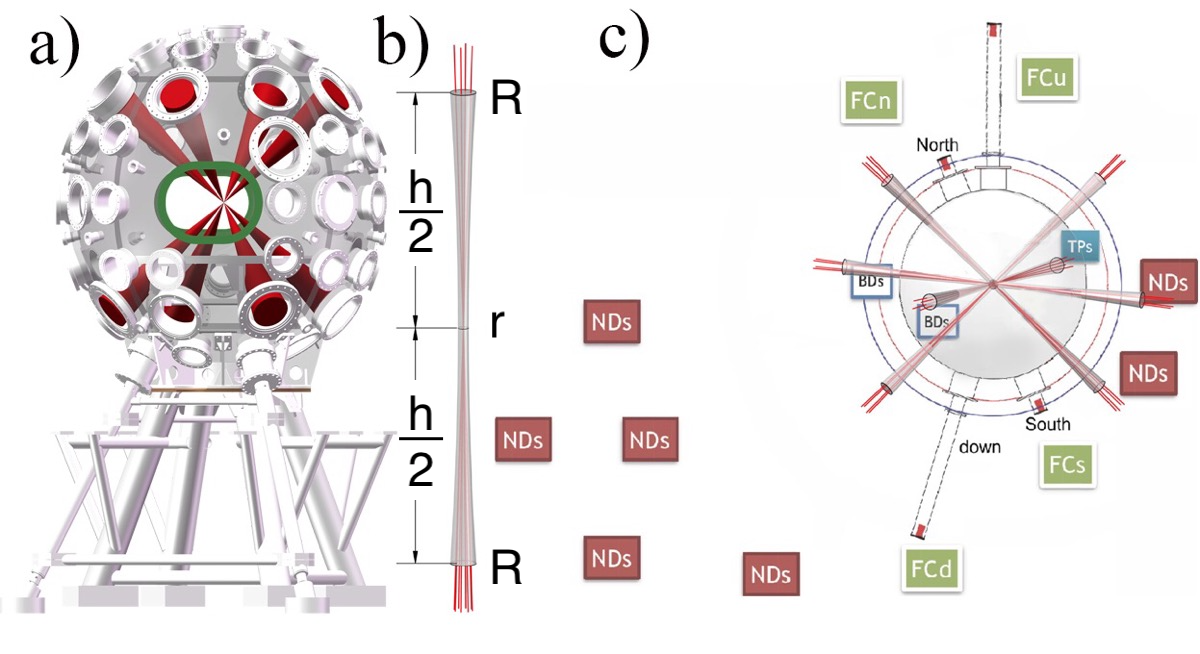}
\caption{
\footnotesize (color online) Experimental setup. a) Schematic view of the target chamber with the 8 laser mirror guides.  b) Schematic view of the target geometry.  c) Schematic view of the chamber with the location of the detectors. The abbreviations NDs is for neutron detectors, FCn(s) for faraday cup north(south), FCd(u) for faraday cup down(up) and TPs for Thomson parabolas.
}
\label{fig1:Setup}
\end{figure}

At the SGII-Up laser facility\cite{Zhao2015}, 8 beam lasers can be used to obtain exactly what we discussed above. A schematic view of the target chamber with the 8 laser mirror guides is given in fig.\ref{fig1:Setup}a). Using 4 beams impinging on a flat target from the up (u), and 4 laser beams impinging from the down (d), we can obtain a beam of ions moving from u to d, and another beam of ions moving in the opposite direction (d to t) \cite{Csernai2019}. In fig.\ref{fig1:Setup}b, we display a schematic view of the target geometry where the laser beams hit the target surface of radius R. The laser beams have an opening angle around ${8.02^\circ}$ from the lens to the target as shown in Fig 1a and Fig 1b. The laser is focalized at the center of the CD$_2$ target with a radius r. For very thin targets, one gets 2R $\approx$ 2r $\approx$ 150  $\mu$m, presently the highest possible focalization at SGII-Up. The thickness h (as well as r and R) was varied to optimize the number of fusions and to prove that effectively beam-beam collisions are occurring. For h$\rightarrow\infty$, we expect the laser beams to act independently, i.e. produce ion beams, which are reflected back.  To find the optimal thickness for which the ion beams `follow' the laser direction, we broke the cylindrical symmetry by using 3 laser beams on t and 4 beams on b for some shots. The target thickness was varied from 1mm to 3.6 $\mu$m and the focalization varied from 150 $\mu$m to 400 $\mu$m. The number and energy of the produced ions were measured using faraday cups (FC) located in the t and b directions at 2.9m distance, see figure \ref{fig1:Setup}c). Other FCs were located at different angles in the same plane of the target, which we indicate as north-south plane. Thomson parabolas (TP) with image plates (IP) at the focal plane. Bubble detectors (BD) were used to measure the number of fusions close to the target. Liquid and plastic neutron detectors (ND) were located outside the scattering chamber at different distances to measure neutrons and they were calibrated on the BD\cite{Mirfayzi2015}. 
A schematic view of the chamber with the location of the detectors is given in figure \ref{fig1:Setup}c. 

The `trick' of using 4 and 3 laser beams respectively demonstrated that even for the thickest target, 1mm, the ion beams follow the laser direction. In fact the FC located in the opposite direction of the 4 laser beams showed higher ion signals. For these `asymmetric' shots, the total laser energy was about 12 kJ and the pulse duration 2 ns. The FC located in the north-south plane gave essentially no signal, demonstrating the high directionality and out of equilibrium nature of the plasma. The TP gave no signal as well, thus suggesting that no ions of kinetic energy above roughly 100 keV (the minimum energy for which the image plate are sensitive) were produced in the north-south plane. SGII-Up facility offers also the possibility to vary the lasers energy and 
the pulse duration by keeping the power constant. Thus we varied the laser energy from about 19 kJ in 3 ns to 2 kJ in 250 ps. Highest compressions were found with the longest pulse duration giving about 10$^8$ neutrons, while the highest plasma ion kinetic energies were found with the shortest pulse duration and lower neutron yields of the order of 10$^5$.

\begin{figure}[h]
\centering
\includegraphics[width=0.475\textwidth]{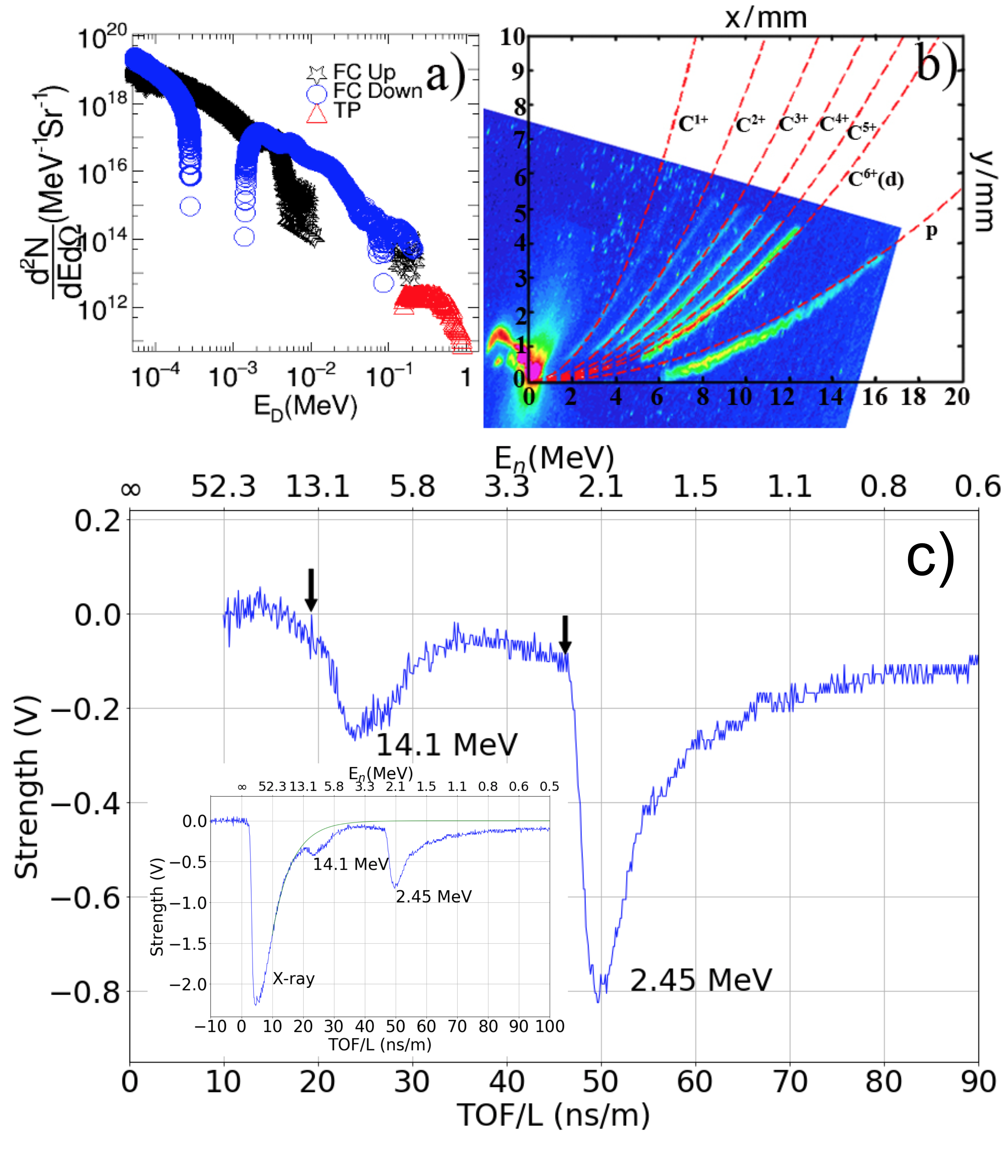}
\caption{\footnotesize (color online)  a) Deuterium energy distribution from the top (full circles) and bottom (open circles) FC, and TP (full triangles). b) TP spectra relative to the laser irradiation of CD$_2$ target with a laser pulse duration 0.5 ns and energy 3.8 kJ.  c) TOF results for 14.1 MeV and 2.45 MeV neutrons from plastic scintillator detector with a laser pulse duration 3.0 ns and energy 18 kJ.  The two arrows indicate the 14.1 MeV and 2.45 MeV neutron energies.
}
\label{fig2:dNdE}
\end{figure}

In figure \ref{fig2:dNdE}, (a) we show the signals obtained from the t and b FC and (b) the typical signals obtained from the TP (b): p, d and the six charge states of C are visible. For this shot, h=3.6  $\mu$m, 2r=150 $\mu$m, and laser pulse duration 0.5 ns. The distribution plotted in fig.2a demonstrates that we obtained kinetic energy of protons (p) (an impurity of the CD$_2$ target) and deuteriums (d) of the order of MeV. For the moment we would like to focus on the fact that long laser pulses, say 3 ns, produce high ion yields with kinetic energies of the order of 1 keV, while short pulses produce $>$100s keV ions with small yield.  

In a CD$_2$ target, fusion reactions might occur if the kinetic energy of the colliding deuterium ions is sufficient to overcome the Coulomb barrier (or tunnel through it with lower probabilities). An optimal energy for this system is of the order of tens to hundreds of keV. Our attempts above changing the lasers pulse duration, had the intent to optimize the resulting ions kinetic energies and maximize the number of fusions. In the plasma, the main fusion channel reactions are
\begin{eqnarray}
 d+d \rightarrow t+p ~~~Q = 4.03MeV \nonumber \\
 d+d \rightarrow ^{3}He+n ~~~Q = 3.27MeV \nonumber
\label{dd}
\end{eqnarray}
These reactions occur with the same probability, and we refer to these as two body fusion reactions N$_2$. The yield is given by:
\begin{equation}
N_2 = N_i \langle\prod_{dd}\rangle/2
\label{eq2:yield}
\end{equation}
N$_i$=$\rho_0$V is the total number of ions which can be calculated from the CD$_2$ initial density $\rho_0$ and volume V of the target given by the cylinder of thickness h and radius r, fig.\ref{fig1:Setup}b). $\langle\prod_{dd}\rangle$ is the average of eq.(\ref{eq1:prod}), and it depends on the ion kinetic energy distribution. If the plasma is in thermal equilibrium, the factor 2 is needed for identical ions. The energetic neutrons (2.45 MeV) can be measured using the ND. 
The produced t and $^3$He have kinetic energy slightly below 1 MeV and the probability of fusion for such energies ($\sigma(dt)$=0.4b=4e$^{-25}cm$$^2$) is quite large and well known from the literature. Thus we can have ternary fusion processes i.e.:
\begin{eqnarray}
 t+d \rightarrow \alpha+n ~~~Q = 17.59 MeV \nonumber \\
 ^{3}He+d \rightarrow \alpha+p ~~~Q = 18.35 MeV \nonumber
\label{td}
\end{eqnarray}
This is the onset of nuclear catastrophic reactions, i.e. reactions that can release large energy in the plasma and warm it up. The number of fusions N$_3$ can be easily calculated as above:
\begin{equation}
N_3 = N_2 \prod_{dt} = N_2(1-e^{-\Lambda \rho \sigma(dt)}) 
\label{eq3:N3}
\end{equation}
similarly for the other reaction. Notice that in this case we have no average sign, in fact the t($^3$He) has 1.01 MeV kinetic energy and the plasma kinetic energy is relatively low (about 0.55 keV in the best cases, with long laser pulses) and can be neglected. Using the ND we can measure the 14.1 MeV (N$_3$) and the 2.45 MeV neutrons (N$_2$). Since all the quantities entering equation (\ref{eq3:N3}) are known or measured, we can invert the equation and derive the areal density $\Lambda \rho$ shot by shot. Thus with this method we use nuclear reactions to measure the areal density at the time of maximum compression where we expect the ternary yield to be highest. In the scenario described here we got a ratio N$_2$/N$_3$$\approx$5 in some shots revealing a tremendous compression! In these cases the total energy release from ternary fusion reactions is comparable to the two body fusion reactions taking advantage of the large Q-value. This is an extremely important result since further compressions (with more laser energy or other features) might decrease the ratio even more\cite{Bonasera2004}, thus fuel targets might be prepared with smaller concentration of the radioactive tritium for applications \cite{Csernai2019,Csernai2018}. 

An example of the neutron measurement is displayed in figure \ref{fig2:dNdE} c), obtained using a plastic scintillator BC420, with 5cm thick lead shielding, located at L=3.3 m from the target. For this shot, the target thickness h=79  $\mu$m, focalization r=300  $\mu$m, total laser energy 18.7 kJ and pulse duration 3 ns. The raw time of flight (TOF) spectrum (opportunely divided by the detector distance L) is given in the inset. It shows a strong EMP for very short times, followed by a small bump corresponding to the 14.1 MeV neutron and a huge bump at longer times (2.45 MeV). The decay time of the EMP is a characteristic of the detector, associated electronics and can be parameterized as an exponential decay. A `clean' spectrum can be obtained after subtracting the exponential decay and it is plotted in the figure \ref{fig2:dNdE} c). The TOF/L is converted in neutron energy in the top axis. The two neutrons peaks are clearly visible showing that the 14.1 MeV peak is quite comparable to the 2.45 MeV, a signature of the high compression. An important feature is that the first peak seems shifted from the 14.1 MeV birth value suggesting that neutrons collide with other ions before exiting the dense plasma. Thus quaternary collisions are important as well and we will examine them more in detail below.

\begin{figure}[ht]
\centering
\includegraphics[width=0.475\textwidth]{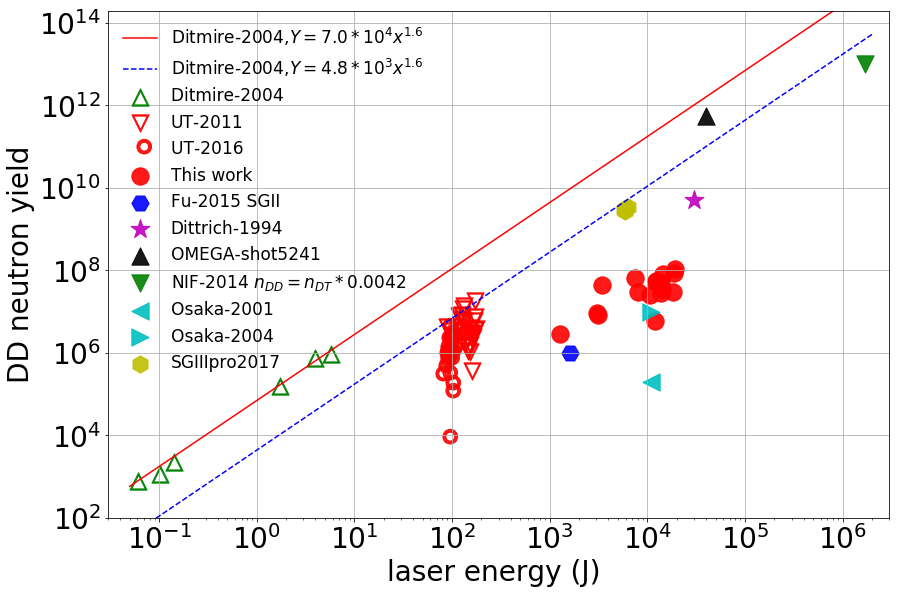}
\caption{\footnotesize (color online) Fusion yield as function of laser energy. Different experimental results Ditmire-2004\cite{Madison2004}, UT-2011\cite{Bang2013}, UT-2016\cite{Quevedo2018}, Fu-2015 SGII\cite{Fu2015}, Dittrich-1994\cite{Dittrich1994} , NIF-2014\cite{Nif2014} , Osaka
-2001\cite{Kodama2001}, Osaka-2004\cite{Kodama2004}, OMEGA-shot5241\cite{Laboratory1995}and SGIIIpro2017\cite{Ren2017} are indicated in the inset.}
\label{fig4:dd}
\end{figure}

In figure \ref{fig4:dd} we plot dd fusion yield as function of laser energy in joules. Our results are given by the full circles and compared to other experiments. The open symbols refer to yields obtained with short pulse lasers and cluster targets\cite{Quevedo2018,Bang2013,Madison2004}. We have corrected the parameterization given in\cite{Madison2004} (full line), to take into account the deuterium concentration (dashed line). It is in good agreement with ref.\cite{Krása2014} which studied fusions in thick CD$_2$ targets using a ps-laser. We have not been able to find in the literature experimental neutron yields above the full line in figure \ref{fig4:dd}. 

The measured neutron spectra clearly suggest that they interact on their way out from the dense plasma. They can collide elastically with the hydrogen contained in the target, and, depending on its energy, also inelastically with C (possibly breaking it into 3$\alpha$ through the Hoyle or higher C exited states) and D atoms. Even though the shift in energy cannot be precisely asserted because of the detector response, we can estimate its effects.  The number of quaternary reactions N$_4$ can be obtained from eq.(\ref{eq3:N3}) substituting 4 with 3 and changing the cross sections depending on the reaction channel. We are interested in the neutron energy depletion from its birth value and it can be estimated from the difference N'$_3$=N$_3$-N$_4$:
\begin{equation}
 N'_3 = N_3 - N_4 = N_3(1-1+e^{-\Lambda \rho \sigma(nC;nD;...)})
\label{eq4}
\end{equation}

\begin{figure}[ht]
\centering
\includegraphics[width=0.475\textwidth]{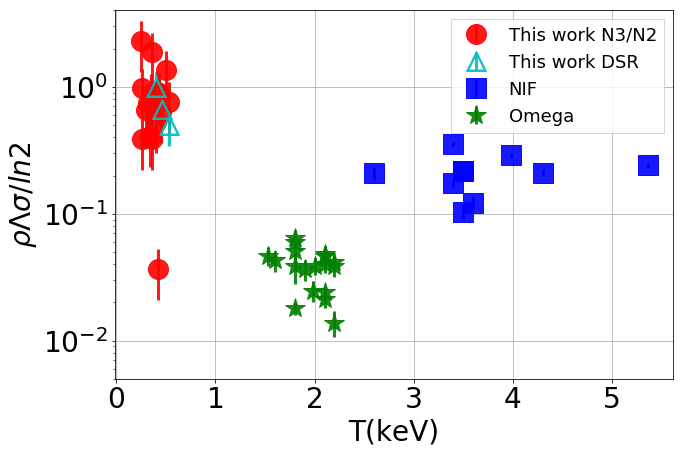}
\caption{\footnotesize (color online) $\Lambda \rho \sigma$/ln2 obtained from eq.(\ref{eq4}) vs T from eq.(\ref{eq1:prod}). Omega and NIF data are derived from the experiments\cite{Casey_phd2012}, using the Down Scatter Ratio\cite{LePape_prl2018,Hurricane2014}. Our results using the DSR method ($N_4$/$N_3$) are given by the open triangle symbols in good agreement with the $N_3$/$N_2$ ratios.
}
\label{fig5}
\end{figure}

In optimal conditions to warm up the plasma, we need the neutrons to release most of their kinetic energy, thus N'$_3<$N$_4$ which after some simple algebra leads to the condition:  $\Lambda \rho \sigma$(nC;nD;..)$>$ln2. The reaction cross-section of n with C, D, H and other ions are known. We can define an average reaction cross-section over the C and D content of our CD$_2$ target taking into account their concentrations. For the 14.1 MeV, we obtained a total cross section, $\sigma_{nCD_{2}}$ = $\sigma_{nC}$ +2$\sigma_{nD}$ = 1.5 b+ 2$\times$0.85 b =3.2 b.  From the quaternary collisions we can also obtain an estimate of the areal density using the Down Scatter Ratio (DSR), the ratio of the number of neutrons with energy between 10 and 12 MeV ($N_4$) divided by the total number of ternary fusion reactions ($N_3$-14.1 MeV neutrons): $\rho\Lambda$=(20.4$\pm0.6$)DSR \cite{LePape_prl2018,Hurricane2014,Johnson2012,Nif2014}. 

The temperature can be obtained assuming, for sake of comparison to other data, that at the time when neutrons are mostly produced (at the highest compression) the plasma is in equilibrium. From the number of two body fusions and the ratio of ternary to two body fusions we can obtain the average dd fusion cross section, see the appendix.  In figure \ref{fig5} we plot the quantity $\Lambda \rho \sigma$/ln2 obtained from eq.(\ref{eq4}) vs T from eq.(\ref{eq1:prod})  and  eq.(\ref{eq2:yield}). If this quantity is much larger than one then most of the 14.1 MeV neutrons have been depleted. In the same plot we display some results obtained at NIF (squares) and at Omega (stars), using 0.9b average cross-section ($\sigma_{nT}$=1.0b and $\sigma_{nD}=$0.85b).
Our results reach values around and above one, apart one shot obtained with the smallest laser energy and pulse duration. The NIF results\cite{Casey_phd2012} are located below one while Omega\cite{Casey_phd2012} is well below such value. It seems that the price to pay is the smaller T in our case. Of course a comparison with the other experiments cannot be done if thermal equilibrium or not is definitely asserted. The DSR method agrees with the ratio $N_3$/$N_2$ for the cases where statistics is large enough to estimate the number of quaternary reactions, open triangles in Figure \ref{fig5}.

In conclusion, in this paper we demonstrated the reaching of record areal high densities in laser compressed plasmas adopting cylindrical symmetry. The ratio of the 14.1 MeV and the 2.45 MeV neutrons gives a direct information of the areal density $\Lambda \rho_{D}$\cite{Kang2008,Blue1981}, confirmed by the DSR method mostly in use at the Omega and the NIF laboratories with DT targets. If we further assume that the range is equal to the thickness of the target at maximum compression, similar to the UT results\cite{Zhang2017, Quevedo2018}, we can write $\Lambda\cong$ N$_i^{1/3}/\rho_{D}^{1/3}$ apart a coefficient of the order of one. The maximum observed areal density in our shots was $\Lambda \rho_D$=4.8 $\pm$ 1.5 g/cm$^2$, N$_i\approx$2.3 $\times$10$^{18}$ thus $\rho_D$=2.0 $\pm$ 0.9 kg/cm$^{3}$,  using the relation between $\Lambda$ and $\rho_D$. The ions kinetic energy is about 0.6 keV from which we can obtain an average velocity $v=\sqrt{2T/m_d}$; a characteristic time $\Lambda$/v=52 $\pm$ 9 ps can be associated with the plasma lifetime at maximum compression or the beam-beam crossing time. It can be compared with the bang time obtained at NIF\cite{Hurricane2014} of the order of 150 ps. Longer plasma lifetimes imply more fusion reactions and this must be balanced with the higher densities obtained with our geometry. An optimal determination of all these factors may optimize the efficiency of nuclear fuel burning.

\section*{Appendix}

\begin{figure}[ht]
\centering
\includegraphics[width=0.475\textwidth]{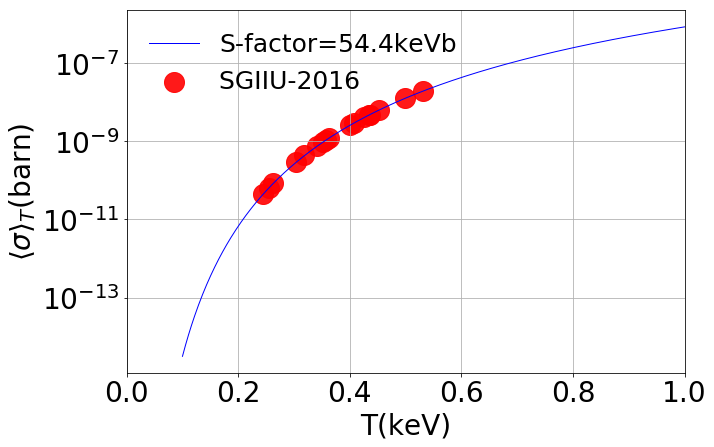}
\caption{\footnotesize (color online) The average cross section as function of temperature with the Maxwell-Boltzmann distribution, expressed by eq. (\ref{eq:eq_S}) . The red points are the experimental cross section data from eq.(\ref{eq:cross_section}).
}
\label{fig6}
\end{figure}

From eq.(\ref{eq1:prod}) and eq.(\ref{eq2:yield}), we can derive 
\begin{equation}
\langle\sigma\rangle_{T}=\frac{-ln\left(1-\frac{N_2}{2N_i}\right)}{\Lambda\rho}.
\label{eq:cross_section}
\end{equation}

Since N$_2$ (from ND), $\Lambda \rho$ (from the ratio N$_3$/N$_2$) and N$_i$ (from geometry) are known, we can derive $\langle\sigma\rangle_{T}$ for each shot. Assuming that the cross-section is not modified by the plasma environment, we can derive the corresponding T\cite{Burbidge1957,Bang2013} if we assume the plasma to be in equilibrium or the C.M. energy of the colliding ions if it is out of equilibrium\cite{Zheng2013}. In the case of beam-beam collisions, half of the ions travel in one direction and half in the opposite one, thus a further factor of 2 is needed in eq.(\ref{eq2:yield}) and the average is over very narrow ion distributions in energy and angle as in our case. Since we are going to use this result for the purpose of comparison to other experiments, we derive an effective T from $\langle\sigma\rangle_T$ obtained inverting eq.(\ref{eq1:prod}) (the x-axis in figure \ref{fig5}),  as expressed by eq.(\ref{eq:eq_S}) and shown in figure \ref{fig6}.
Using the steepest descent method, one can get the average cross section as function of temperature from the Maxwell-Boltzmann distribution \cite{Barbarino_phd2015}, 
\begin{equation}
\langle\sigma \rangle_{T}=\frac{4}{\sqrt3}\frac{S}{T}  e^{-\frac{3E_{G}}{T}}.
\label{eq:eq_S}
\end{equation}
For $d+d \rightarrow {^{3}He}+n$ reaction, a constant S-factor S = 54.4 keVb is set and $E_{G} = (\frac{T\sqrt{b}}{2})^{\frac{2}{3}}$ is the Gamow energy.
The ratio N$_2$/N$_3$ is obtained by integrating the neutron signal from 14.1 MeV to 2.5 MeV (N$_3$) and below 2.5 MeV up to 1.0MeV for N$_2$. The largest error comes from the minimum energy adopted to estimate the 2.45 MeV. This is because 14.1 MeV and 2.45 MeV might be shifted down below 2.45 MeV. Furthermore, some neutrons might be bounce back neutrons of any energy depending on their trajectory and detector distances. Thus we estimated the 2.45 MeV by integrating the signal down to 1 MeV in one case and to 0.25 MeV in another case. We averaged the results and estimated the resulting error in about 30\%. 
\section*{Acknowledgements}
This work was partially supported by the Strategic Priority Research Program of the Chinese Academy of Sciences (Grant No. XDB16) and Doctoral Scientific Research Foundation of Inner Mongolia University for Nationalities(No. BS365, BS400), the National Natural Science Foundation of China under Contract No. 11421505, the US Department of Energy-NNSA DE-NA0003841 (CENTAUR) and the Key Research Program of Frontier Sciences of the CAS under Grant No. QYZDJ-SSW-SLH002. AB thanks the CAS President's International Fellowship Initiative No. 2015VWA070 and the Inner Mongolia University of Nationalities for the warm hospitality and support during his stay in China while this work was completed. We thank X.H. Yuan from Shanghai Jiao Tong University, J.J. He from National Astronomical Observatories, Chinese Academy of Sciences for the neutron detectors and helpful discussions on the experiment. We thank Q.N. Li, H.Y. Zheng, Y.H. Ma, W. Liu, X. Wang from Shanghai Institute Applied Physics, Chinese Academy of Sciences for the help on the CD$_2$ target preparation and the component measurement.



\end{document}